# Estimation of Blood Flow Parameters in the Left Atrial Appendage from 4DCT Dynamic Contrast Enhancement

Lauren M. Severance, Andrew M. Kahn, Juan C. del Alamo, Elliot R. McVeigh

*Abstract*—Cardiac CT is often used clinically in electrophysiology to evaluate cardiac morphology. One such case is to evaluate patients with Atrial Fibrillation (AF). AF can cause regions of slow blood flow and blood stasis through the left atrial appendage (LAA), and therefore, it may be preferable to evaluate blood flow through the LAA in addition to morphology. Although CT cannot measure flow directly, CT data has been used to estimate flow using modeling approaches such as Computational Fluid Dynamics, which take into account the cardiac geometry to simulate flow.

Advances in CT technology now enable high-resolution imaging of the whole heart with low radiation doses. With multi-heartbeat imaging during a contrast injection, we can obtain 4-dimentional CT (4DCT) images to measure dynamic contrast enhancement directly. In this study, we use high-resolution 4DCT to acquire images of contrast enhancement across the LAA over multiple heartbeats. The CT contrast signal at each voxel over time is used to create dynamic contrast enhancement maps of parameters derived from a gamma-variate fit. These contrast enhancement maps enable quantification and visualization of spatial-temporal characteristics of flow parameters across the LAA.

## I. Introduction

CARDIAC CT has been used for many years in electrophysiology to plan cardiac ablation and evaluate morphology. Modern fast CT techniques give us the opportunity to also evaluate blood flow and blood stasis as a risk marker for thrombus [1]. These studies often use Computational Fluid Dynamics to estimate a blood flow field from the CT data, and one important clinical application is to evaluate patients with Atrial Fibrillation (AF) [2]. AF occurs when the heart's upper chambers beat irregularly, creating regions of slow blood flow and blood stasis through the left atrium (LA) and left atrial appendage (LAA), which is a sac that protrudes from the LA. Blood stasis may promote formation of a thrombus in the LAA, which can travel to the brain and cause a stroke. As a result, stroke risk is five times higher in patients with AF than in the general population, and the vast majority of atrial thrombi in patients with AF are located in the LAA [3].

While CT is not able to directly measure flow like transesophageal echocardiography or MRI [4], [5], CT can obtain high-resolution images of the whole heart at low radiation doses. Using multi-heartbeat imaging during a contrast injection, CT can provide high spatial resolution estimates of dynamic contrast agent concentration to directly evaluate blood flow and blood stasis. The purpose of this project was to create accurate local estimates of contrast agent concentration over time, as well as estimates of blood flow parameters, at high spatial resolution over the entire LAA.

## II. Methods

### A. Image Acquisition

4-dimensional CT (4DCT) images were acquired using a custom CT imaging protocol similar to baseline cardiac CT perfusion [6]. The contrast injection consisted of 80 ml of Iopamido, Isovue 370, Bracco, at 5 ml/sec with a 50 ml saline flush. Acquisitions were made with a cardiac gated CT "half-scan", 280 ms gantry rotation on the GE Revolution CT 256 slice scanner using a low-dose variant of the "dynamic perfusion protocol" acquisition. 26 128-slice volumes were collected every 1-3 heartbeats over 40-60 seconds, depending on patient heart rate, and a 27th volume was collected approximately 100 seconds after the first volume. Volumes were collected at 100 kVp and 100 mA. All timeframes were acquired using prospective gating during a 170 ms window around end-systole (35% R-R).

### B. Image Reconstruction and Postprocessing

Images were reconstructed with a 512x512x128 matrix size. Voxel size was ~0.5 mm in the xy-plane and 1.25 mm in z. Images were reconstructed using the vendor provided deep learning based "DLIR-high" method. After reconstruction, all timeframes were aligned using nonrigid registration for motion

Research reported in this publication was supported by the NIH/NHLBI grant R01HL160024.

L.M.S. and E.R.M. are with the Shu Chien – Gene Lay Department of Bioengineering, University of California, San Diego, La Jolla, CA 92093 USA (email: lseveran@ucsd.edu, emcveigh@ucsd.edu).

A.M.K. and E.R.M. are with the Division of Cardiovascular Medicine, University of California San Diego, La Jolla, CA 92093 USA (akahn@health.ucsd.edu).

E.R.M. is also with the Department of Radiology, University of California, San Diego, La Jolla, CA 92093 USA.

J.C.A. is with the Mechanical Engineering Department and the Department of Cardiology, University of Washington, Seattle, WA 98195 USA (email: juancar@uw.edu).

correction [7]. Multiplanar reconstruction was used to cut a slice through the LAA, producing a series of 27 2-dimensional images over a region of interest surrounding the LAA. Prior to fitting with the gamma-variate model, each timeframe was smoothed using a gaussian filter with a standard deviation of 1 mm, and voxels with a cumulative sum across time <100 Hounsfield units (HU) were excluded from fitting. A summary of the pipeline for image postprocessing and the generation of dynamic contrast enhancement maps is provided in Fig. 1.

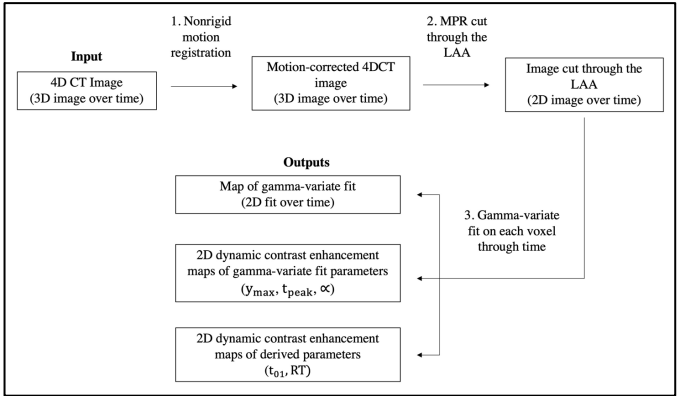

Fig. 1. Pipeline to generate dynamic contrast enhancement maps. The 4-dimensional CT (4DCT) image consists of 27 3D volumes, each collected during a 170 ms window around end-systole. In Step 1, the 27 timeframes are aligned using a nonrigid registration motion correction to produce registered 4DCT images. In Step 2, multiplanar reconstruction is used to create a 2-dimensional (2D) slice through the LAA, creating a set of 27 2D images of the LAA. In Step 3, the 1-dimensional signal of each pixel over time is fit using the gamma-variate model. After each pixel is fit, the parameters produced by the model can be used to create a map of the gamma-variate fit intensity over time (2D fit over time), dynamic contrast enhancement maps of parameters of the fit ($y_{max}$, $t_{peak}$, $\propto$), and dynamic contrast enhancement maps of derived parameters ($t_{01}$, RT).

### C. Blood Transit Model Overview

Blood flow through the LAA was quantified using the gamma-variate model, which has previously been used to describe a bolus as it passes through a series of compartments [8], [9] and is widely accepted for fitting CT contrast concentration as a function of time [10], [11]. To apply the model to our 4DCT data, we used the gamma-variate formulation of Bindschadler et al [11] and Madsen et al [9]:

$$y(t) = y_{max} \left(\frac{t}{t_{peak}}\right)^{\propto} e^{\alpha\left(1-\frac{t}{t_{peak}}\right)} - y_b \quad (1)$$

where $y(t)$ is the CT image intensity as a function of time, $y_{max}$, $t_{peak}$, and $\propto$ are parameters of the fit, $y_b$ is the measured CT signal baseline, and t is time.

### D. Gamma-variate Model Fit and Calculation of Flow Parameters

For each image voxel, we fit the 1-dimensional (1D) signal across time using the gamma-variate equation to obtain a voxel-specific model of the 3 gamma-variate parameters ($y_{max}$, $t_{peak}$, and $\propto$). Then, using this voxel-specific model, we derived additional parameters to describe the blood flow, $t_{01}$ (see Fig. 2) and residence time (RT) (see Eq. 2).

The CT image contrast $y(t)$ for each voxel was derived as the 1D voxel signal over time minus a baseline signal ($y_b$). The baseline signal was calculated as the mean of timepoints 1-3 in the descending aorta.

The 1D signal of CT image contrast over time was fit to the gamma-variate model using the MATLAB "fit" function (MathWorks Inc., Natick, Massachusetts, R2022a). Table 1 shows the lower and upper limits defined for each fit parameter.

TABLE I: Gamma-variate fit parameters limits

| Parameter | Lower limit | Upper limit |
|---|---|---|
| $\propto$ | 0 | Infinity |
| $y_{max}$ | Maximum data value of CT image contrast - 40 HU | Maximum data value of CT image contrast + 40 HU |
| $t_{peak}$ | Time of maximum data contrast minus 5 seconds | Time of maximum data contrast plus 5 seconds |

After obtaining a voxel-specific model of the gamma-variate fit, $t_{01}$ and RT were derived. $t_{01}$ was calculated as the time at which the gamma-variate fit reaches 1% of its maximum value. RT was calculated according to the following equation:

$$RT = \frac{\int_0^\infty t * y(t) dt}{\int_0^\infty y(t) dt} \quad (2)$$

where $y(t)$ is the analytical model for CT image contrast as a function of time, resulting from the gamma variate fit, and t is time. Fig. 2 shows the signal contrast over time, the corresponding gamma-variate model fit, and several model parameters for an example image voxel.

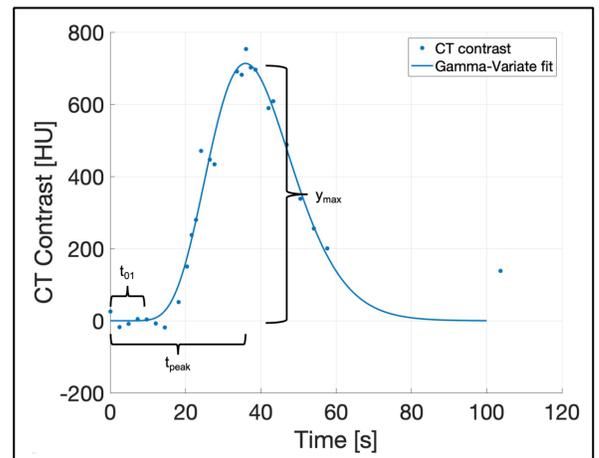

Fig. 2. Example gamma-variate fit and model parameters. CT signal contrast as a function of time (blue dots) is fit using the gamma-variate model (blue line). CT contrast is calculated as the CT image intensity at a given pixel minus a baseline signal. The baseline signal is calculated as the average CT signal at the first 3 timepoints in a region of interest in the descending aorta. The gamma-variate fit parameters $t_{peak}$ and $y_{max}$ are direct outputs of the model. $t_{01}$ is a derived parameter and is defined as the time at which the signal reaches 1% of its peak value.

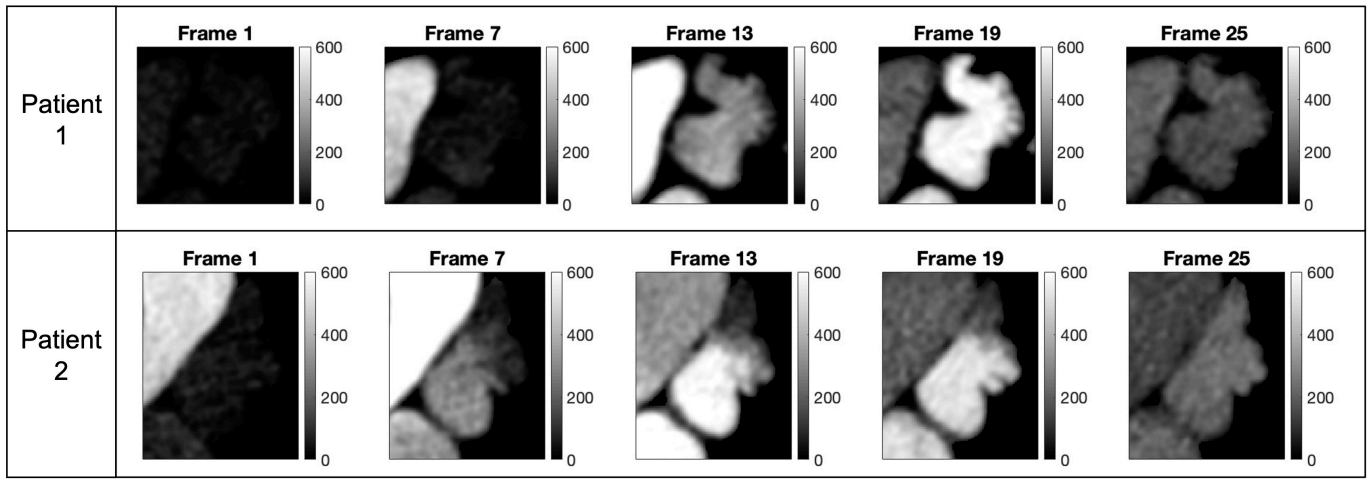

Fig 3. CT images of contrast enhancement over time for 2 patients. The post-processed images enable detailed visualization of contrast enhancement over time, and the trajectory of contrast enhancement can be evaluated across the LAA. For each patient, 5 example timeframes are shown from a dataset containing 27 timeframes.

## III. Results

### A. Contrast Enhancement Images

High-resolution CT images were acquired at 27 timeframes. After motion registration and multiplanar reconstruction, the images were used to visualize contrast enhancement across the LAA over time. These images were then used as input to the gamma-variate model to calculate parameters of the flow at each image voxel. Fig. 3 shows the contrast enhancement images for 2 patients at 5 example timeframes. In frame 1, the LAA appears dark due to the lack of contrast. Over the course of imaging, the signal within much of the LAA increases up to a maximum intensity, as the contrast enters the LAA. After reaching a maximum, the signal decreases as contrast flows out of the LAA. Across 17 patients in the clinic who underwent imaging with this unique protocol, the mean dose length product (DLP) was 8.28±1.05 mSv.

### B. Dynamic Contrast Enhancement Maps

Parameters derived from the gamma-variate fit were used to create 2D dynamic contrast enhancement maps. Each dynamic contrast enhancement map provides a spatially smooth map of a model parameter and can be used to visualize blood flow patterns across the LAA. Fig 4. shows the dynamic contrast enhancement map of $RT$ for two clinical cases. In Patient 1, the $RT$ calculated from the gamma-variate fit is similar for pixels in the proximal, middle, and distal regions of the LAA, and the map of $RT$ appears uniform. In Patient 2, the contrast enhancement of pixels in the middle and distal regions is delayed and reaches a lower peak than pixels in the proximal region, resulting in increased $RT$ in the middle and distal regions. The map of $RT$ across the LAA for Patient 2 displays a delayed filling pattern.

## IV. Discussion

We have evaluated parameters of the blood flow across the LAA using dynamic CT contrast enhancement. With fast, high-resolution CT, we can create detailed enhancement maps across the LAA to characterize parameters of the flow. In this study, we found it was sufficient to use the data collected at 27 timepoints at a relatively low dose (approximately 100 mA x-ray fluence) to create smooth enhancement maps. Although the signal at a given voxel may be noisy over time, fitting with the gamma-variate model provides a smooth estimate of the contrast enhancement over time.

Among patients with AF who underwent imaging with this protocol in our clinic, we have observed a large dynamic range of $RT$ and $y_{max}$ values across the LAA with highly variable spatial patterns. It is clear that a significant number of patients who are in sustained AF have virtually no difference in flow parameters between the LA and the distal LAA; this indicates these patients are likely not at higher risk for thrombus generation in the LAA.

In a small cohort of patients, the mean DLP was comparable to other cardiac exams, including cardiac CT angiography. To further limit radiation exposure, future studies are warranted to evaluate the feasibility of using fewer timeframes to create dynamic contrast enhancement maps. The exact number and timing of images needed to obtain reasonable estimates of fit parameters remains an open question. We are currently developing a machine learning network that will allow us to compute dynamic contrast enhancement of maps the full 3D volume of the LA and LAA in a reasonable amount of time. Future optimizations will include the number of images required and the sampling frequency, minimum z-coverage required to capture the LAA during breathing, optimal injection profile for contrast agent for this application, and optimal spatial resolution/patient dose trade-off.

## Conclusion

Using a new dynamic contrast CT protocol, we have created a technique to look at blood flow dynamics in the LAA in patients with AF. These data provide a potential method to evaluate the risk of thrombus formation in the LAA and probability of stroke.

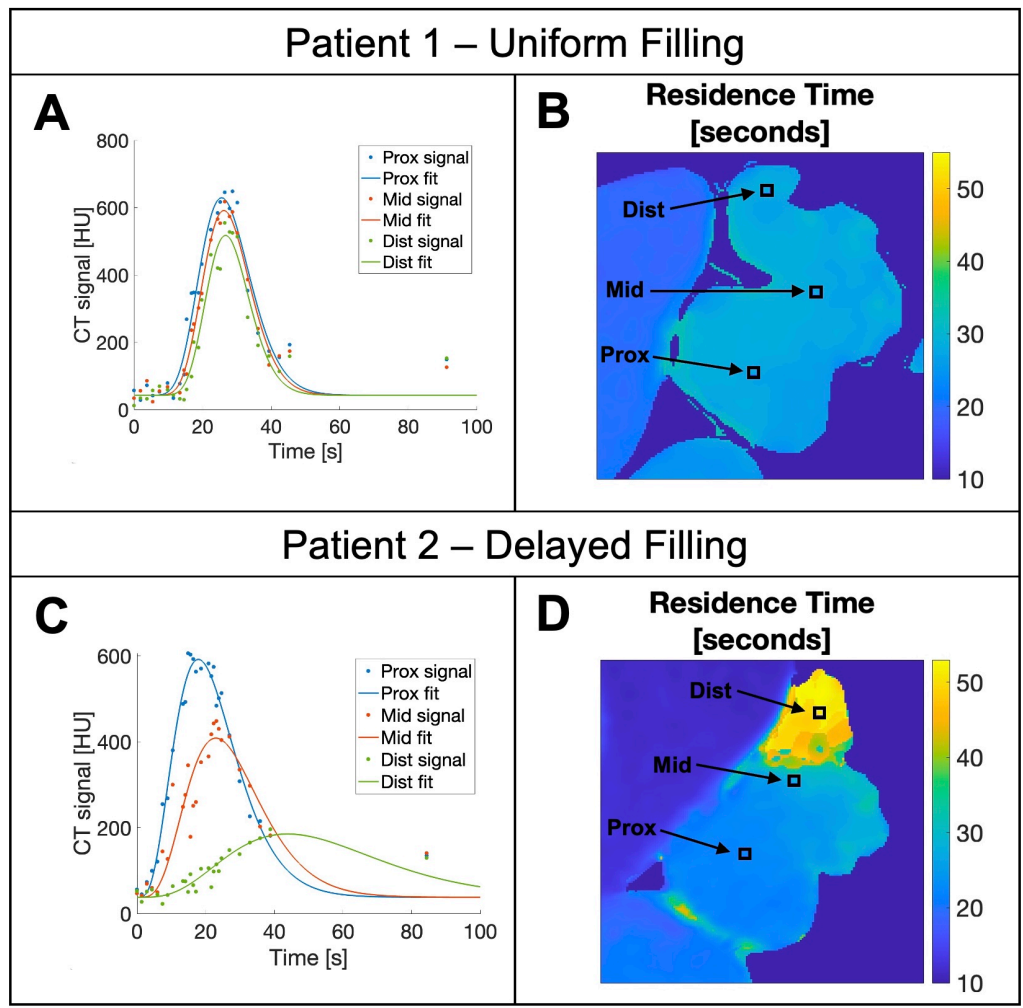

Fig. 4. Maps of residence time (RT) generated from the gamma-variate fit for a slice through the left atrial appendage (LAA). (A and C) CT signal intensity over time and corresponding gamma-variate fit for a single voxel in the proximal region (Prox signal, Prox fit), middle region (Mid signal, Mid fit), and distal region (Dist signal, Dist fit) of the left atrial appendage (LAA). (B and D) The gamma-variate fit from each pixel can be used to create a smooth map, or dynamic contrast enhancement map, of each flow parameter, including blood residence time (RT). In some patients, (Patient 1), the gamma-variate fits of voxels in the proximal, middle, and distal LAA overlap, and the dynamic contrast enhancement map of RT appears uniform across the LAA (B). In other patients, (Patient 2), the gamma-variate fit in voxels of the proximal, middle, and distal LAA does not overlap (C), and the dynamic contrast enhancement map of RT indicates delayed filling in the middle and distal regions of the LAA (D).